\documentclass[letter]{aa} % for the letters

\usepackage{graphicx}
\usepackage[varg]{txfonts}
\usepackage{natbib}
\usepackage{hyperref} %To add links in your PDF file, use the package "hyperref" with options according to your LaTeX or PDFLaTeX drivers.

\defcitealias{Boley_etal_2012}{B12}
\begin{document}
\title{Constraints on the gas content of the Fomalhaut debris belt\thanks{Based on \emph{Herschel} observations. \emph{Herschel} is an ESA space observatory with science instruments provided by European-led Principal Investigator consortia and with important participation from NASA.}}
\subtitle{Can gas-dust interactions explain the belt's morphology?}
\author{G. Cataldi\inst{1,2} \and A. Brandeker\inst{1,2} \and G. Olofsson\inst{1,2} \and C.~H. Chen\inst{3} \and W.~R.~F. Dent\inst{4} \and I. Kamp\inst{5} \and A. Roberge\inst{6} \and B. Vandenbussche\inst{7}}
\institute{AlbaNova University Centre, Stockholm University, Department of Astronomy, SE-106 91 Stockholm, Sweden \and
Stockholm University Astrobiology Centre, SE-106 91 Stockholm, Sweden \and
Space Telescope Science Institute, 3700 San Martin Drive, Baltimore, MD 21218, USA \and
ALMA SCO, Alonso de Cordova 3107, Vitacura, Casilla 763 0355, Santiago, Chile \and
Kapteyn Astronomical Institute, University of Groningen, Postbus 800, 9700 AV Groningen, The Netherlands \and
Goddard Center for Astrobiology, Goddard Space Flight Center, Greenbelt, MD, 20771, USA \and
Institute of Astronomy KU Leuven, Celestijnenlaan 200D, B-3001 Leuven, Belgium}
\date{Received; accepted}

\abstract
 % context heading (optional), leave it empty if necessary
{The 440\,Myr old main-sequence A-star Fomalhaut is surrounded by an eccentric debris belt with sharp edges. This sort of a morphology is usually attributed to planetary perturbations, but the orbit of the only planetary candidate detected so far, Fomalhaut~b, is too eccentric to efficiently shape the belt. Alternative models that could account for the morphology without invoking a planet are stellar encounters and gas-dust interactions.}
 % aims heading (mandatory)
{We aim to test the possibility of gas-dust interactions as the origin of the observed morphology by putting upper limits on the total gas content of the Fomalhaut belt.}
 % methods heading (mandatory)
{We derive upper limits on the \ion{C}{ii} 158\,$\mu$m and \ion{O}{i} 63\,$\mu$m emission by using non-detections from the Photodetector Array Camera and Spectrometer (PACS) onboard the \textit{Herschel Space Observatory}. Line fluxes are converted into total gas mass using the non-local thermodynamic equilibrium (non-LTE) code \textsc{radex}. We consider two different cases for the elemental abundances of the gas: solar abundances and abundances similar to those observed for the gas in the $\beta$~Pictoris debris disc.}
 % results heading (mandatory)
{The gas mass is shown to be below the millimetre dust mass by a factor of at least $\sim$3 (for solar abundances) respectively $\sim$300 (for $\beta$~Pic-like abundances).}
 % conclusions heading (optional)
{The lack of gas co-spatial with the dust implies that gas-dust interactions cannot efficiently shape the Fomalhaut debris belt. The morphology is therefore more likely due to a yet unseen planet (Fomalhaut~c) or stellar encounters.}

\keywords{circumstellar matter -- planetary systems -- Stars: individual: Fomalhaut -- Methods: observational -- Hydrodynamics -- Infrared: general}

\maketitle

%%%%%%%%%%%%%%%%%%%%%%%%%%%%%%%%%%%%%%%%%%%%%%%%%%%%%%%%%%%%%%%%%%%%%%%

\section{Introduction}
Fomalhaut is one of the best studied examples out of several hundred main-sequence stars known to be surrounded by dusty discs commonly known as \textit{debris discs}. The presence of circumstellar material around this nearby \citep[ 7.7\,pc;][]{vanLeeuwen_2007}, $440\pm40$\,Myr old \citep{Mamajek_2012} A3V star was first inferred by the detection of an infrared excess above the stellar photosphere due to thermal emission from micron-sized dust grains \citep{Aumann_1985}. The dusty debris belt around Fomalhaut has been resolved at infrared wavelengths with \textit{Spitzer} \citep{Stapelfeldt_etal_2004} and the \textit{Herschel} Photodetector Array Camera and Spectrometer (PACS) \citep{Acke_etal_2012}. The dust is thought to be derived from continuous collisions of larger planetesimals or cometary objects \citep{Backman_Paresce_1993}. To trace these parent bodies, observations at longer wavelengths sensitive to millimetre-sized grains are used. Millimetre-sized grains are less affected by radiation pressure and thus more closely follow the distribution of the parent bodies. \citet{Holland_etal_1998} were the first to resolve the Fomalhaut belt in the sub-millimetre. Recent data from the Atacama Large Millimeter/submillimeter Array (ALMA) show a remarkably narrow belt with a semi-major axis $a\sim140$\,AU \citep[][hereafter \citetalias{Boley_etal_2012}]{Boley_etal_2012}.

The belt has also been observed in scattered star light with the \textit{Hubble Space Telescope} \citep{Kalas_etal_2005,Kalas_etal_2013}. Two key morphological characteristics emerging from the \citet{Kalas_etal_2005} observations are i) the ring's eccentricity of $e\sim0.1$ and ii) its sharp inner edge. Both features have been seen as evidence for a planet orbiting the star just inside the belt \citep[e.g.][]{Quillen_2006}. A candidate planetary body (Fomalhaut~b) was detected \citep{Kalas_etal_2008}, but subsequent observations showed that its orbit is highly eccentric ($e=0.8\pm0.1$), making it unlikely to be responsible for the observed morphology \citep{Kalas_etal_2013,Beust_etal_2014,Tamayo_2014}. There remains the possibility that a different, hereto unseen planet is shaping the belt; infrared surveys were only able to exclude planets with masses larger than a Jupiter mass \citep{Kalas_etal_2008,Marengo_etal_2009,Janson_etal_2012,Janson_etal_2014}. Actually, the extreme orbit of Fomalhaut~b might be a natural consequence of the presence of an additional planet in a moderately eccentric orbit \citep{Faramaz_etal_2015}. Alternatively, the observed belt eccentricity may be caused by stellar encounters \citep[e.g.\ ][]{Larwood_Kalas_2001,Jalali_Tremaine_2012}. In particular, Fomalhaut is part of a wide triple system. \citet{Shannon_etal_2014} recently showed that secular interactions or close encounters with one of the companions could result in the observed belt eccentricity.

We focus on yet another candidate mechanism to explain the observed morphology. It is known that gas-dust interactions can result in a clumping instability, organising the dust into narrow rings \citep{Klahr_Lin_2005,Besla_Wu_2007}. \citet{Lyra_Kuchner_2013} recently presented the first 2D simulations of this instability and found that some rings develop small eccentricities, resulting in a morphology similar to that observed for the Fomalhaut debris belt. The mechanism obviously requires the presence of gas beside the dust (typically, a dust-to-gas ratio of $\epsilon\lesssim1$ is required), but we know from objects such as $\beta$ Pictoris \citep[e.g.][]{Olofsson_etal_2001} or 49~Ceti \citep[e.g.][]{Hughes_etal_2008} that debris discs are not always devoid of gas. We test the applicability of the clumping instability to the case of Fomalhaut by searching for \ion{C}{ii} 158\,$\mu$m and \ion{O}{i} 63\,$\mu$m gas emission using \textit{Herschel} PACS. We do not detect any of the emission lines and use our data to put stringent upper limits on the gas content of the Fomalhaut debris belt.

\section{Observations and data reduction}
\begin{figure}
\centering
\includegraphics[width=\hsize]{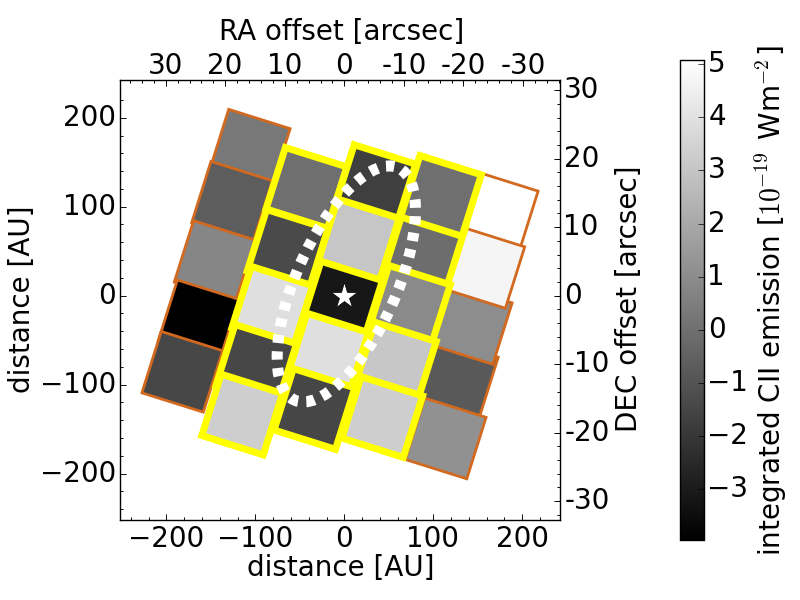}
\caption{Placement of the 25 PACS spaxels relative to Fomalhaut (marked by the white star) in the case of the \ion{C}{ii} observations. The grey scale indicates the integrated \ion{C}{ii} emission for each spaxel within a window of width equal to 2.5*FWHM of the line. The ellipse shows the location of the peak density of mm grains according to the best-fit model by \citetalias{Boley_etal_2012}. The corresponding figure for the \ion{O}{I} data looks very similar. The 15 spaxels with a yellow bold edge were summed to produce Fig.\ \ref{upperlimits}.}
\label{PACS_CII_lineemission}
\end{figure}

Fomalhaut was observed using PACS \citep{Poglitsch_etal_2010} onboard the \textit{Herschel Space Observatory} \citep{Pilbratt_etal_2010}. The integral field unit PACS consists of 25 spatial pixels (spaxels), each covering $9.4\arcsec\times9.4\arcsec$ on the sky. We used PACS in line spectroscopy mode (PacsLineSpec) to observe the \ion{C}{ii} 158\,$\mu$m and the \ion{O}{i} 63\,$\mu$m line regions (observation IDs 1342257220 and 1342210402 respectively). We observed in chop/nod mode. Figure \ref{PACS_CII_lineemission} shows the placement of the PACS spaxels relative to the Fomalhaut debris belt in the case of the \ion{C}{ii} observations. We reduced the data using the ``background normalisation'' pipeline script within the \textit{Herschel} Interactive Processing Environment (HIPE) version 12.0 \citep{Ott_2010}. The background normalisation pipeline is recommended for faint sources or long observations since it corrects more efficiently for detector drifts compared to the ``calibration block'' pipeline. The pipeline performs bad pixel flagging, cosmic ray detection, chop on/off subtraction, spectral flat fielding, and re-binning. We choose a re-binning with oversample=4 (thus gaining spectral resolution\footnote{With this choice of re-binning, the resolution is $\sim$60\,km\,s$^{-1}$ for the \ion{C}{II} data and $\sim$22\,km\,s$^{-1}$ for the \ion{O}{I} data.} by making use of the redundancy in the data) and upsample=1 (thus keeping the individual data points independent). Finally, the pipeline averages the two nodding positions and calibrates the flux. The noise in the PACS spectra is lowest at the centre of the spectral range and increases versus the spectral edges. We discard the very noisy spectral edges. We then subtract the continuum by fitting linear polynomials to the spectra, where the line region is masked. The pipeline generated noise estimate ``stddev'' is used to give the highest weight to the data points at the centre of the spectral range\footnote{Even if the noise estimate delivered by the pipeline probably underestimates the true uncertainty, it is still useful for fitting the continuum, since it describes the relative change of the noise over the spectral range.}.
%see http://herschel.esac.esa.int/hcss-doc-12.0/index.jsp#pacs_spec:pacs_spec for a discussion of oversample, upsample and why the data edges can safely be rejected

\section{Results and analysis}
%In the following sections, we describe the derivation of upper limits on the \ion{C}{ii} and \ion{O}{i} line emission and how this information is used to constrain the total gas content of the Fomalhaut debris belt.

\subsection{Upper limits on the \ion{C}{ii} and \ion{O}{i} emission from PACS}\label{flux_upperlimits}
Whereas both lines remain undetected, we do detect the dust continuum at both wavelengths in the spaxels covering the belt. At 63\,$\mu$m, we see additional continuum emission at the position of the star. Within the absolute flux calibration accuracy of PACS, the total detected continuum is consistent with the photometry presented by \citet{Acke_etal_2012}.

We adopt a forward modelling strategy to compute upper limits on the line flux. We simulate PACS observations from a map of the line intensity on the sky. The PACS Spectrometer beams (version 3)\footnote{\url{http://herschel.esac.esa.int/twiki/bin/view/Public/PacsCalibrationWeb\#PACS\_spectrometer\_calibration}} are used to calculate the integrated line flux registered in each spaxel from the input map. The line is assumed to be unresolved. Thus the computed flux is assumed to be contained in a Gaussian-shaped line with a full width at half maximum (FWHM) determined by the PACS instrument\footnote{0.126\,$\mu$m for the \ion{C}{ii} 158\,$\mu$m line and 0.018\,$\mu$m for the \ion{O}{i} 63\,$\mu$m line; see section 4.7.1 of the PACS Observer's Manual, \url{http://herschel.esac.esa.int/Docs/PACS/pdf/pacs_om.pdf}}. These simulated spectra can then be compared to the real observations. The input sky map is constructed under the assumption that the gas is co-spatial with the dust, since we are interested in constraining the possibility of gas-dust interactions. We take the best-fit model of the distribution of mm grains derived by \citetalias{Boley_etal_2012} as a description of the line luminosity density (energy from line emission per volume) and derive a sky map assuming optically thin emission. The only free parameter in the model is a global, positive scaling of the sky map, determining the total flux received from the belt. We adopt a simple Bayesian approach with a flat prior for the scaling factor. Assuming Gaussian noise for the PACS data points, a posterior probability distribution can be derived, which is used to compute an upper limit on the scaling factor and thus on the total line emission. The noise is estimated for each spaxel individually from two 2.5*FWHM wide spectral windows, placed sufficiently far from the line centre so to avoid any potential line emission. However, the windows are also sufficiently far from the noisy edges of the spectral range. Thus we do not significantly overestimate the noise in the line region itself.

\begin{figure}
\centering
\includegraphics[width=\hsize]{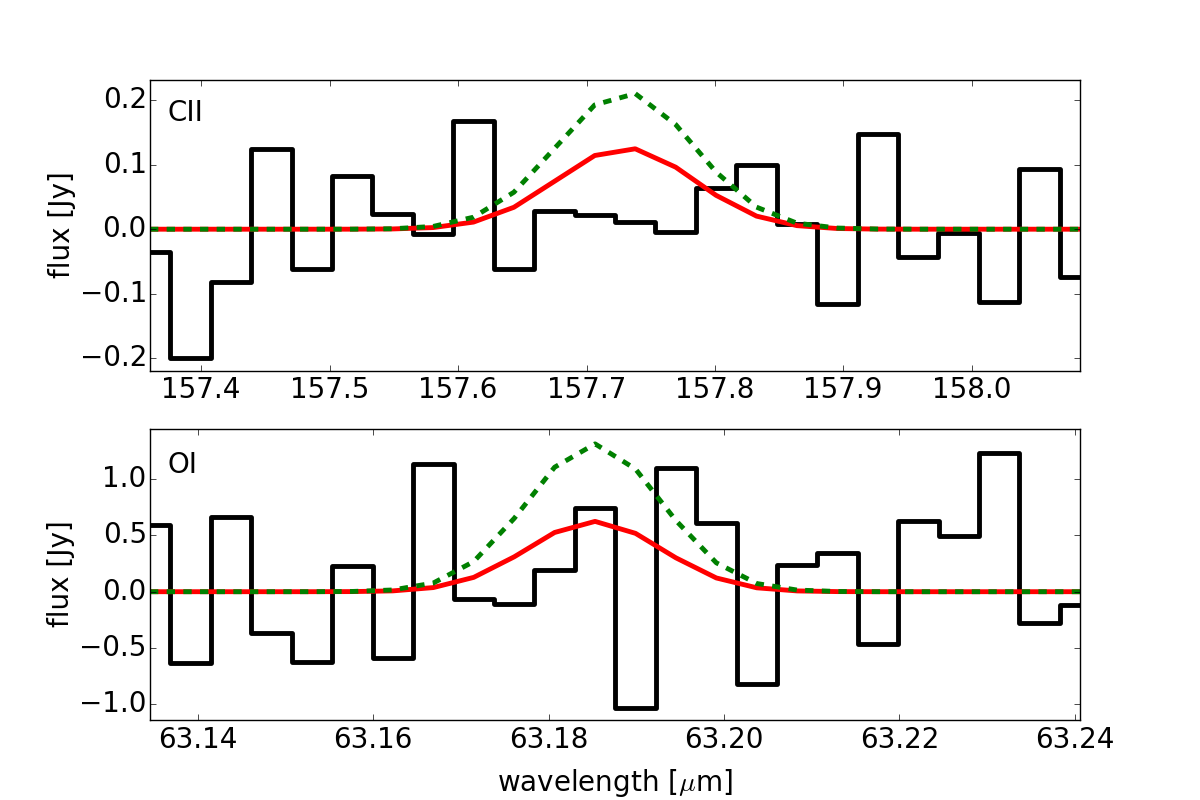}
\caption{Line fluxes (\ion{C}{ii} and \ion{O}{i}) obtained when co-adding spaxels marked with bold edges in Fig.\ \ref{PACS_CII_lineemission}. A 99\% confidence level upper limit was derived by integrating over a 2*FWHM wide wavelength range centred at the line wavelength and propagating the error on the integrated flux. The dashed line shows the Gaussian profile containing this upper limit flux. The plain line shows the profile obtained by co-adding the same spaxels containing our simulated spectra corresponding to the Bayesian upper limit, demonstrating the stronger limit obtained with our detailed model.}
\label{upperlimits}
\end{figure}

Table \ref{upper_limits_flux_table} shows upper limits on the total line emission of gas co-spatial with the dust at 99\% confidence level. The \ion{C}{II} line gives a significantly stronger upper limit because of the lower noise level in these data. We also estimated upper limits using Monte Carlo simulations by repeatedly fitting our model to new realisations of the data with added noise. This approach gives upper limits that are smaller by $\sim$5\% for \ion{C}{ii} and $\sim$20\% for \ion{O}{i} compared to the Bayesian calculation. We state the more conservative Bayesian values here. Compared to co-adding flux from the spaxels covering the belt (Fig.\ \ref{PACS_CII_lineemission}), our approach using a model of the spatial distribution allows for a stronger upper limit (Fig.\ \ref{upperlimits}). This is expected since more information is used to constrain the maximum flux compatible with the data.

\begin{table}
\caption{Upper limits (99\% confidence level) on line fluxes.} % title of Table
\label{upper_limits_flux_table} % is used to refer this table in the text
\centering % used for centering table
\begin{tabular}{c c} % centered columns
\hline\hline % inserts double horizontal lines
Line & Flux \\    % table heading
 & (W\,m$^{-2}$) \\
\hline % inserts single horizontal line
\ion{C}{ii} 158\,$\mu$m & $<2.2\times10^{-18}$ \\   % inserting body of the table
\ion{O}{i} 63\,$\mu$m&  $<1.0\times10^{-17}$\\
\hline  %inserts single line
\end{tabular}
\end{table}

\subsection{Upper limits on the total gas mass in the Fomalhaut debris belt}
We use the non-local thermodynamic equilibrium (non-LTE) excitation and radiative transfer code \textsc{radex} \citep{vanderTak_etal_2007} to convert the upper limits on the \ion{C}{ii} and \ion{O}{i} line emission (Table \ref{upper_limits_flux_table}) into gas masses. The emission lines are assumed to be excited by collisions with atomic hydrogen and electrons. We do not attempt to calculate the thermal balance, but instead consider a range of kinetic gas temperatures between 48\,K (the dust temperature, \citetalias{Boley_etal_2012}) and $10^4$\,K, which is the temperature where collisional ionisation of both \element{C} and \element{O} becomes important. The \element{C}/\element{O} ratio is fixed to solar. For the abundance of the other elements, we consider two cases: solar abundances \citep{Lodders_2003} and abundances as observed in the $\beta$ Pictoris gaseous debris disc, where the gas mass is dominated by \element{C} and \element{O} \citep{Cataldi_etal_2014,Brandeker_etal_2015_subm}. The latter choice of abundances is further motivated by a recent study of the 49~Ceti debris disc, suggesting a C-rich gas similar to $\beta$ Pic \citep{Roberge_etal_2014}.

\begin{figure}
\centering
\includegraphics[width=\hsize]{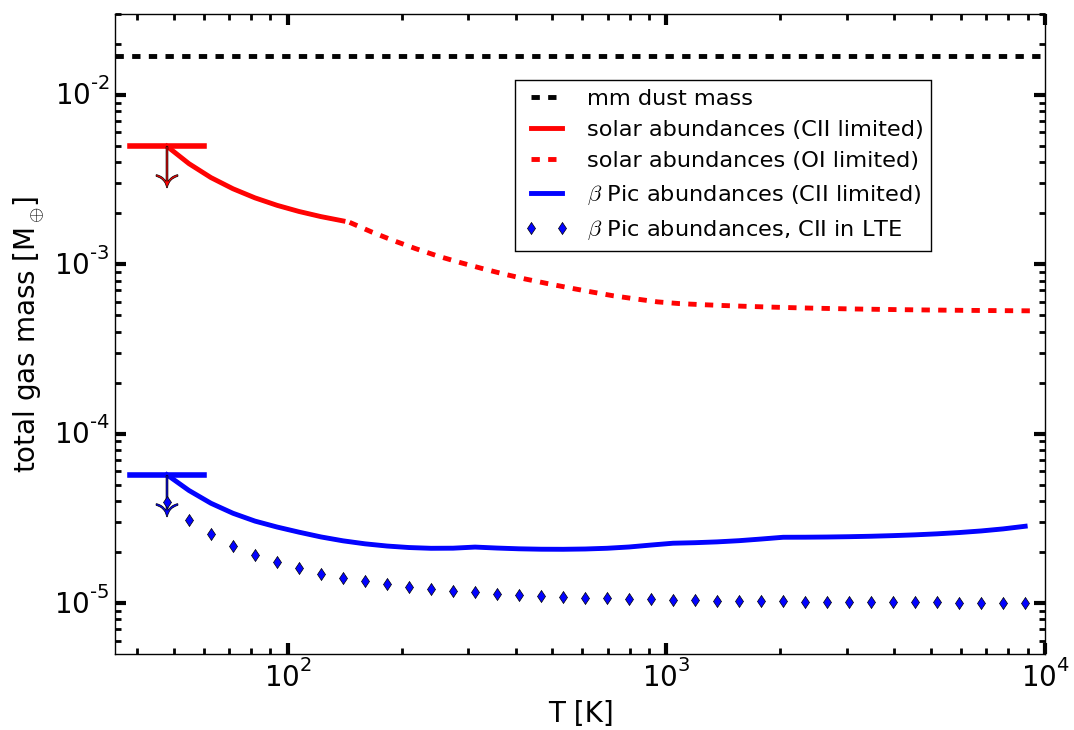}
\caption{Total gas mass as a function of the kinetic temperature, derived from models reproducing either the \ion{C}{ii} or \ion{O}{i} flux limit from Table \ref{upper_limits_flux_table} (as indicated in the legend). The \element{C}/\element{O} ratio is fixed to solar. The gas masses are derived using the non-LTE code \textsc{radex} for both solar and $\beta$~Pic-like elemental abundances. In addition, the gas mass derived from a simple LTE calculation is shown for $\beta$~Pic-like abundances. The total gas mass remains below the mm dust mass (horizontal dashed line) over the entire temperature range. We take the gas masses at the lowest kinetic temperature as our upper limits (indicated by an arrow and shown in Table \ref{upper_limits_gas_table}).}
\label{total_gas_mass}
\end{figure}

For a given temperature, we search the amount of gas that is reproducing either of the upper limits (\ion{C}{ii} or \ion{O}{i}). Usually, the gas mass is limited by the \ion{C}{ii} flux. The \ion{O}{i} flux is the limiting factor only for kinetic temperatures in excess of 150\,K and solar abundances (see Fig.\ \ref{total_gas_mass}). The densities of the collision partners (\element[-][]{e} and \element{H}) determine the excitation of the emission lines and therefore the total amount of gas necessary to reproduce a certain flux, but the total amount of gas also implies certain values for the \element{H} and \element[-][]{e} densities. Therefore, to construct a self-consistent model, an iterative approach is necessary. Starting with a guess of the \element[-][]{e} and \element{H} densities, we use \textsc{radex} to compute the amount of \element[+][]{C} (or \element{O} if \ion{O}{i} is limiting the total gas mass) reproducing the corresponding flux upper limit. Next, the ionisation fraction of \element{C} is estimated. We assume equilibrium between photoionisation and recombination. Ionising photons from two different sources are included: Fomalhaut's photosphere from an ATLAS9 model \citep{Castelli_Kurucz_2004} with $T_\mathrm{eff}=8500$\,K, $\log g=4.0$ and $\log Z=0$, and the interstellar UV field \citep[taken from][]{Draine_1978}, where the latter is the dominant component. Ionisation by cosmic rays is negligible. A simplifying assumption is that all electrons are coming from the ionisation of \element{C}, i.e.\ $n_{\element{C}}=n_{\element[-][]{e}}$. This should be an excellent approximation if the gas has $\beta$~Pic-like abundances. For solar abundances, we are likely underestimating the electron density because of the presence of highly ionised elements such as \element{Mg} or \element{Fe}. This only affects the ionisation calculation, but not the excitation of the lines, since for solar abundances, \element{H} is dominating the collisional excitation\footnote{In the case of the \element{H}-poor $\beta$~Pic-like abundances, collisional excitation is completely dominated by \element[-][]{e}.}. The ionisation fraction of \element{C} is giving us the total amount of \element{C} and \element[-][]{e} in the model. The total gas mass (and in particular the \element{H} density) can then be derived from the assumed elemental abundances, where elements up to atomic number $Z=38$ are considered (hydrogen through strontium). We iterate until the \element{H} and \element[-][]{e} input densities are equal to the output densities implied by the total gas mass. We assume \element{O} to be completely neutral in our models because of its high ionisation potential. This is justified by computing the ionisation of \element{O} for various electron densities and temperatures, showing that \element{O} remains largely neutral for $n_\mathrm{e}\gtrsim10^{-2}$\,cm$^{-3}$.

While \textsc{radex} assumes a homogeneous medium, we calculated the upper limits on the line emission using the \citetalias{Boley_etal_2012} dust profile as a description of the line luminosity density. We convert a gas mass output from \textsc{radex} to a density by distributing the mass according to the \citetalias{Boley_etal_2012} profile and taking the peak density. For example, the \element[+][]{C} mass from \textsc{radex} implies an \element[-][]{e} density via the peak density of the \citetalias{Boley_etal_2012} density profile. Changing this mass-to-density conversion by e.g.\ taking only half the peak density has only a minor effect on our results.

Table \ref{upper_limits_gas_table} shows the upper limits on the gas mass and corresponding lower limits on the dust-to-gas ratio derived for the two elemental abundances considered. We also list upper limits on the \element{C} and \element{O} mass, the range of collision partner densities occurring and the maximum column densities, showing that the emission is optically thin. Figure \ref{total_gas_mass} shows the gas mass as a function of the kinetic temperature. The upper limit on the gas mass comes from the lowest kinetic temperature (T=48\,K) and is below the millimetre dust mass derived by \citetalias{Boley_etal_2012}. For illustration, we also performed a simple calculation of the amount of \element{C} necessary to reproduce the upper limit on the \ion{C}{ii} flux in the case of LTE for $\beta$~Pic-like abundances. It is seen that LTE is not a very accurate approximation, demonstrating the need for a non-LTE code like \textsc{radex} for the type of environment investigated here.

\begin{table*}
\caption{Upper limits on the total gas mass (and corresponding lower limits on the gas-to-dust ratio), the \element{C} mass and the \element{O} mass. The gas masses are calculated from 99\% confidence level upper limits on the \ion{C}{ii} and \ion{O}{i} flux. We also list the range of collision partner densities and maximum column densities occurring in the models.} % title of Table
\label{upper_limits_gas_table} % is used to refer this table in the text
\centering % used for centering table
\begin{tabular}{c c c c c c c c c c c} % centered columns
\hline\hline % inserts double horizontal lines
abundances & $M_{\mathrm{gas}}$& $\epsilon$\tablefootmark{a} &$M_{\element{C}}$&$M_{\element{O}}$& \multicolumn{2}{c}{\element[-][]{e} density} & \multicolumn{2}{c}{\element{H} density} & $N_\mathrm{max}$\tablefootmark{d} (\element[+][]{C}) & $N_\mathrm{max}$\tablefootmark{d} (\element{O}) \\
& &&& & min & max & min & max & & \\
 &  ($\mathrm{M}_\oplus$) & & ($\mathrm{M}_\oplus$) & ($\mathrm{M}_\oplus$) & \multicolumn{2}{c}{(cm$^{-3}$)} & \multicolumn{2}{c}{(cm$^{-3}$)}& (cm$^{-2}$) & (cm$^{-2}$)  \\
\hline
solar\tablefootmark{b} & $<5.0\times10^{-3}$ & $>3.4$ &$<1.2\times10^{-5}$&$<3.2\times10^{-5}$& 2 & 10 & $5\times10^{3}$ & $5\times10^{4}$ & $8\times10^{14}$ & $2\times10^{15}$ \\
$\beta$ Pic\tablefootmark{c} & $<5.7\times10^{-5}$ &>298 &$<1.6\times10^{-5}$&$<4.1\times10^{-5}$& 6 & 12 & $8\times10^{-2}$ & $2\times10^{-1}$ & $1\times10^{15}$ & $3\times10^{15}$\\
\hline  %inserts single line
\end{tabular}
\tablefoot{
\tablefoottext{a}{Dust-to-gas ratio, using the mm dust mass $M_\mathrm{mm}=0.017$\,$\mathrm{M}_\oplus$ derived by \citetalias{Boley_etal_2012}.}
\tablefoottext{b}{Assuming solar elemental abundances.}
\tablefoottext{c}{Assuming elemental abundances similar to the $\beta$~Pic gas disc.}
\tablefoottext{d}{Assuming a typical line width of $\sim$2\,km/s, the corresponding optical depths are below $10^{-2}$.}
}
\end{table*}

\section{Summary and conclusion}
Gas-dust interactions have been shown to concentrate dust into narrow, eccentric rings via a clumping instability. A necessary condition for the instability to develop is a dust-to-gas ratio $\epsilon\lesssim1$ \citep{Lyra_Kuchner_2013}, although the instability is expected to be maximised for $\epsilon\approx0.2$ (W.~Lyra 2014, private communication). We know from e.g.\ $\beta$~Pictoris or 49~Ceti that some debris discs contain gas such that $\epsilon$ is indeed of the order of unity. Prior to the present work, the only information on the gas content of the Fomalhaut debris belt came from non-detections of \element{CO}. \citet{Dent_etal_1995} used a non-detection of the J=3-2 transition to put upper limits on the \element{CO} gas mass assuming LTE. Recently, \citet{Matra_etal_2014} used an ALMA non-detection of the same transition to place a $3\sigma$ upper limit of $\sim$$5\times 10^{-4}$\,M$_\oplus$ on the CO content of the Fomalhaut belt (i.e.\ a dust-to-CO ratio $\gtrsim$30), taking the crucial importance of non-LTE effects into account. However, since \element{CO} is expected to be photo-dissociated by stellar or interstellar UV photons on a short timescale compared to the lifetime of the system \citep{Kamp_Bertoldi_2000,Visser_etal_2009}, the absence of \element{CO} does not strongly constrain the total gas mass. Constraints on the atomic gas mass are more useful. We derived upper limits on the atomic gas mass using non-detections of \ion{C}{ii} and \ion{O}{i} emission and assuming either solar abundances or abundances as those observed in the $\beta$~Pic debris disc. The total gas mass is shown to be smaller than the dust mass. Because of the low gas content, gas-dust interactions are not expected to be working efficiently in the Fomalhaut belt. Therefore, the data indirectly suggest a second, yet unseen planet Fomalhaut~c or stellar encounters as the cause for the morphology of the Fomalhaut debris belt.

\begin{acknowledgements}
We would like to thank the referee, Jane Greaves, for useful and constructive comments that helped to clarify this manuscript. We also thank Elena Puga from the \textit{Herschel} Science Centre Helpdesk for support with the \textit{Herschel} beam products. This research has made use of the SIMBAD database (operated at CDS, Strasbourg, France), the NIST Atomic Spectra Database, the NORAD-Atomic-Data database and NASA's Astrophysics Data System. PACS has been developed by a consortium of institutes led by MPE (Germany) and including UVIE (Austria); KU Leuven, CSL, IMEC (Belgium); CEA, LAM (France); MPIA (Germany); INAF-IFSI/OAA/OAP/OAT, LENS, SISSA (Italy); IAC (Spain). This development has been supported by the funding agencies BMVIT (Austria), ESA-PRODEX (Belgium), CEA/CNES (France), DLR (Germany), ASI/INAF (Italy), and CICYT/MCYT (Spain).
\end{acknowledgements}

\bibliographystyle{aa}
\bibliography{bibliography}

\end{document}